\documentclass[aps,prb,amsmath,amssymb,reprint,authoryear,numbers,sort&compress,superscriptaddress]{revtex4-2}
\usepackage{color}
\usepackage{graphicx}% Include  files
\usepackage{subfigure}
\usepackage{dcolumn}% Align table columns on decimal point
\usepackage{bm}% bold math
\usepackage[T1]{fontenc}
\usepackage{newtxtext, newtxmath}
\usepackage{multirow}
\begin{document}

\title{High-temperature thermoelectric properties with Th$_{3-x}$Te$_4$}% \\

\author{Jizhu Hu}
\affiliation{
	Center for Phononics and Thermal Energy Science, School of Physics Science and Engineering, Tongji University, Shanghai 200092, China}
\affiliation{
	Max Plank Institute for Polymer Research, Ackermannweg 10, 55128 Mainz, Germany}
\author{Jinxin Zhong}
\affiliation{
	Center for Phononics and Thermal Energy Science, School of Physics Science and Engineering, Tongji University, Shanghai 200092, China}	
%\author{Jie Chen}
%\affiliation{
%	Center for Phononics and Thermal Energy Science, School of Physics Science and Engineering, Tongji University, Shanghai 200092, China}
%\author{George Fytas}
%\affiliation{
%	Max Plank Institute for Polymer Research, Ackermannweg 10, 55128 Mainz, Germany}
\author{Jun Zhou}
\email{zhoujunzhou@njnu.edu.cn}
\affiliation{
NNU-SULI Thermal Energy Research Center and Center for Quantum Transport and Thermal Energy Science, School of Physics and Technology, Nanjing Normal University, Nanjing, Jiangsu 210023, China}

\date{\today}

\begin{abstract}
Th$_3$Te$_4$ materials are potential candidates for commercial thermoelectric (TE) materials at high-temperature due to their superior physical properties.
We incorporate the multiband Boltzmann transport equations with firstprinciples
calculations to theoretically investigate the TE properties of Th$_3$Te$_4$ materials.
As a demonstration of our method, the TE properties of La$_3$Te$_4$ are similar with that of Ce$_3$Te$_4$ at low-temperature, which is consistent with the experiment. Then
we systematically calculate the electrical conductivity, the Seebeck coefficient, and the power factor of the two materials above based on parameters obtained from first-principles calculations as well as several other fitting parameters.
Our results reveal that for the electron--optical-phonon scattering at high temperatures, a linear dependence of optical phonon energy on temperature explains better the experimental results than a constant optical phonon energy.
Based on this, we predict that the TE properties of Ce$_3$Te$_4$ is better than La$_3$Te$_4$ at high temperatures and the optimal carrier concentration corresponding to Ce$_3$Te$_4$ shifts upward with increasing temperature.
The optimal carrier concentration of Ce$_3$Te$_4$ is around $1.6\times10^{21}$cm$^{-3}$ with the peak power factor 13.07 $\mu$Wcm$^{-1}$K$^{-2}$ at $T=1200$K.
\end{abstract}

%\pacs{72.20.Pa,72.10.-d}% PACS, the Physics and Astronomy
                             % Classification Scheme.
%\keywords{thermoelectric properties, nanocomposites, minority carriers blocking, majority carriers low-energy filtering, thermoelectric device}%Use showkeys class option if keyword
                              %display desired
\maketitle

\section{Introduction}

Thermoelectric (TE) materials have drawn great interest as solid-state energy converters which can directly convert heat to electricity and vice versa. \cite{Bell2008,disalvo1999thermoelectric,qin2021power}
The energy conversion efficiency of TE materials is characterized by a dimensionless figure of merit $ZT = \sigma S^{2}T/ \kappa$, where $\sigma$ is the electrical conductivity, $S$ is the Seebeck coefficient, $T$ is the absolute temperature, and $\kappa$ is the thermal conductivity consisting of electronic $\kappa_{c}$ and lattice thermal $\kappa_{l}$ conductivity, $\kappa = \kappa_{c} + \kappa_{l}$. 
A higher cooling or power generation efficiency of TE devices requires larger $ZT$ values.
In the past decades, the $ZT$ of TE materials has remained near 1 because $\sigma$, $S$ and $\kappa_{c}$ are coupled to each other.\cite{jonson1980mott}
It is difficult to improve the TE properties of materials by optimizing one of the parameters alone while keeping the others constant. \cite{rowe2018crc}
A larger power factor, defined as $\sigma S^{2}$, is also required to gain larger output power.

Th$_3$P$_4$ (Th refers to rare earth metals, P refers to sulfur group) has long been of interest due to its superior physical properties, such as superconductivity, mixed valence, strong electronic correlation, magnetic properties, optical properties, and TE properties.\cite{Viennois2013}
Th$_3$Te$_4$ is a cubic crystal structure with the space group $I \overline{4} 3d$. 
The Te atoms are hexa-aligned with the rare earth metal lanthanide system through a twisted octahedron.\cite{May2008}
It can be found by stoichiometry that the compounds with Th$_3$Te$_4$ structure have good electrical properties due to the presence of one extra electron.
At the same time, the presence of vacancies leads to disorder and distortion in the lattice, which enhances phonon scattering and leads to a lower lattice thermal conductivity.\cite{wood1985thermoelectric}

The properties of Th$_3$Te$_4$ have previously been investigated by using solid-state diffusion and melt synthesis methods. \cite{ramsey1965phase}
However, the melt synthesis method leads to inhomogeneous sample chemistry and carrier concentration caused by working temperatures up to $2080\sim2280$K.
May \emph{et al.} in 2008 proposed a mechanical alloying method to prepare La$_{x-3}$Te$_4$.\cite{May2008}
This method can effectively avoid the generation of inhomogeneous grains.
The authors estimated the lattice thermal conductivity of La$_{x-3}$Te$_4$ at 573K as $0.2\sim0.4$Wm$^{-1}$K$^{-1}$ through the free electron Lorentz number.
They also measured a power factor of 1.6Wm$^{-1}$K$^{-2}$ and a $ZT$ value of 1.1 for La$_{x-3}$Te$_4$ at 1273K.
Recently, Pr$_{2.74}$Te$_4$ with a $ZT$ value as high as 1.7 was prepared by Cheikh \emph{et al.} using a mechanical alloying method.\cite{cheikh2018praseodymium}
Ce$_{x-3}$Te$_4$ and La$_{x-3}$Te$_4$ have similar TE properties in the low temperature region. \cite{May2012}
Using the first-principle, Wang \emph{et al.} found that the Ce$_{3}$Te$_4$ structure has a $\delta$-peak with 0.21eV in the density of states near the Fermi surface. \cite{wang2011}
Therefore, they predicted that Ce$_{3}$Te$_4$ has excellent TE properties at high temperatures.
Although the localized $f$ electrons in Ce$_{3}$Te$_4$ make the density of states near the Fermi surface sharp, the Seebeck coefficient is not increased by the presence of $f$ electrons. \cite{Vo2014}
This is also confirmed by experimental measurements.

In this paper, the multiband Boltzmann transport equations (BTE) are used to explore and predict the TE transport properties under the relaxation time approximation (RTA). The parameters such as band gap and effective mass of each band are calculated from first-principles calculations
to solve the BTE. The RTA based on the multiband carrier transport model is also used. \cite{Zhou2010,wang2017theoretical}
In order to demonstrate our method, we study the TE properties of La$_3$Te$_4$ and Ce$_3$Te$_4$.
Based on these results, the optimal carrier concentrations for peak of power factor are predicted for the Ce$_3$Te$_4$ materials at high temperatures.
The TE properties of other Th$_{3-x}$Te$_4$ materials can be studied similarly.

\section{Band Structure and Phonon Spectrum }
\label{Band}
  We employ the Vienna ab initio Simulation Package (VASP), \cite{kresse1996} which is based on the density function theory (DFT) and generalized gradient approximation (GGA) with the Perdew-Burke-Ernzerhof (PBE), \cite{perdew1996,perdew1998perdew} to calculate the band structure and phonon spectrum of Th$_3$Te$_4$ materials. The structure of Th$_3$Te$_4$ were relaxed in cell shape, atom positions and volume. A plane-wave energy cutoff of 650 eV and Monkhorst-Pack $\Gamma$-centered k-point meshes of 9$\times$9$\times$9 were employed. For La$_3$Te$_4$, we consider the effect of spin-orbit coupling (SOC).\cite{May2009} Besides, due to the localization of \emph{f}-electrons in Ce$_3$Te$_4$, the on-site Coulomb interaction must be considered to correct the self-interaction for \emph{f}-electrons.\cite{Vo2014} The total energy is converged to less than 10$^{-9}$ eV/unit. To determine the phonon spectrum, a conventional cell is expanded to a 2$\times$2$\times$2 supercell which contains 224 atoms, which further undergoes a structure relaxation. Hellmann-Feynman forces is calculated in relaxed supercell. Finally, The phonon spectrum of Th$_3$Te$_4$ are obtained via utilizing the phonopy package combined with VASP.

\subsection{La$_3$Te$_4$} 
\begin{figure}[htp]
\centering
\includegraphics[width=6cm]{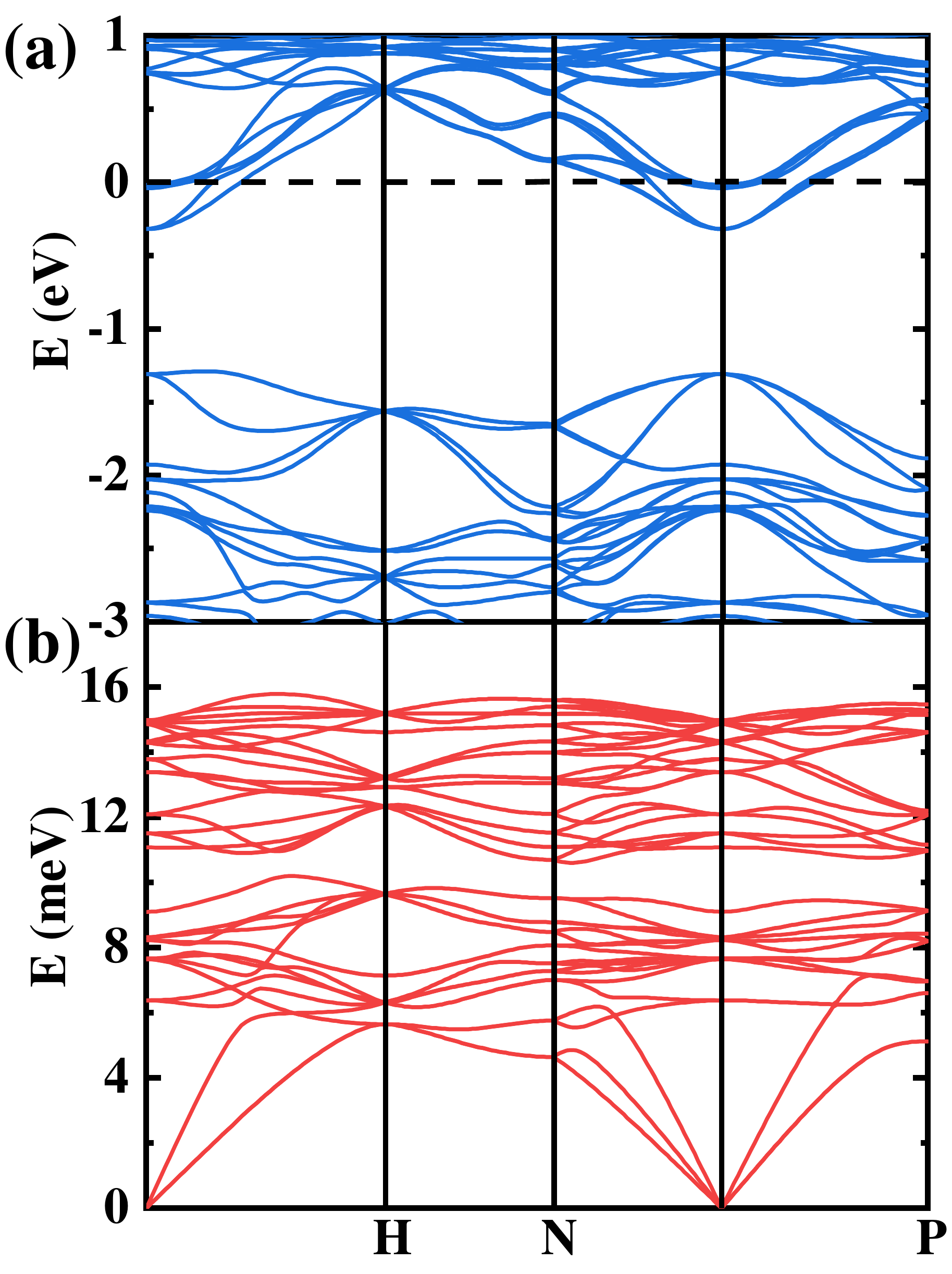}
\caption{(a) The band structure and (b) phonon spectrum of  La$_3$Te$_4$.}
\label{LATE-SP}
\end{figure}

The electronic band structure and phonon spectrum of La$_3$Te$_4$ are shown in Figure \ref{LATE-SP}. The main parameters for calculating the TE properties are summarized in Table \ref{PARA}.

The calculated band structure for La$_3$Te$_4$ is consistent with that in Ref. \onlinecite{perdew1996}, where a direct gap at $\Gamma$ (0.99 eV) was obtained. The energy minima of these bands which relative to \emph{E}$_F$ are \emph{E}$_{min,1}$=-0.316 eV, \emph{E}$_{min,2}$=-0.039 eV, and \emph{E}$_{min,3}$=-0.018 eV, respectively. The effective mass of each conduction and valance bands near the $\Gamma$ is obtained by fitting the band structure. The density-of-state effective mass of each carriers pocket can be expressed as
\begin{eqnarray}
	m^{\ast}_{i,j}=(m^{\ast 2}_{i,j,\|}m^{\ast}_{i,j,\bot})^{\frac{1}{3}},
\end{eqnarray} 
where $m^{\ast}_{i,j,\|}$ ($m^{\ast}_{i,j,\bot})$ is the parallel (perpendicular) effective mass near the band edge and $k_{i,j,\|}$ ($k_{i,j,\bot}$) is the corresponding wave vector. \emph{i} represents the index of electron band. $j$ is the type of carrier, ${j=e}$ for electron and ${j=h}$ for hole. Effective mass and corresponding degeneracy for each conduction and valance bands are shown in Table \ref{EFF}, where spin degeneracy is not included. \par 
Figure \ref{LATE-SP}(b) shows phonon dispersion curves of La$_3$Te$_4$, there are 42 different types of vibration modes in the primitive unit cell, including 3 acoustic modes and 39 optical modes. The longitudinal ($\upsilon_{LA}$) and transverse ($\upsilon_{TA}$) speed of sound can be obtained via fitting acoustic modes at $\Gamma$ point. The low-energy peak (optical mode A$_2$) is 9.09 meV, in agreement with Ref.\,\onlinecite{Viennois2013}.

  \begin{table*}
	\caption{Parameters obtained from band structure and phonon spectrum of La$_3$Te$_4$ and Ce$_3$Te$_4$, such as lattice constant, band gap, energy minima, A$_2$ mode and speed of sound.}
	\begin{tabular}{@{}lcccccccc}
		\toprule
	Materials & $a$(\AA) & $E_g$ & $E_{min,1}$(eV) &$E_{min,2}$(eV) &	$E_{min,3}$(eV) &$\textsc{A}_2$(meV)&$\upsilon_{LA}$(m/s)& $\upsilon_{TA}$(m/s)\\
		\colrule
		La$_3$Te$_4$ & 9.688&0.99&-0.316&-0.039&-0.018&9.09&3357&1989\\
		Ce$_3$Te$_4$ & 9.542&1.07&-0.388&-0.170&-0.005&9.65&3463&2013\\
		\botrule
	\end{tabular}
	\label{PARA}
\end{table*}

\begin{table}
	\caption{Effective mass and corresponding degeneracy for each conduction band and valance band in La$_3$Te$_4$ and Ce$_3$Te$_4$.}
	\begin{tabular}{@{}lccccccc}
		\toprule
	    ~&(i, j) & $m^{\ast}_{i,j,\|}$($m_0$) & $m^{\ast}_{i,j,\bot}$ ($m_0$) & $m^{\ast}_{i,j}$ ($m_0$) & Degeneracy $N_i$\\
	    \colrule
		\multirow{5}*{La$_3$Te$_4$} ~&(1,e) & 0.404&0.3616&0.389 & 2 \\
		~&(2,e) &1.1108&0.9601&1.058 & 1\\
		~&(3,e) &1.1987&1.2481&1.215 & 2\\
		~&(1,h) &0.3412&0.4174&0.341 & 1\\
 	    ~&(2,h) &1.965 &0.9189&1.184 & 1\\ 
		\colrule 
		\multirow{5}*{Ce$_3$Te$_4$} ~&(1,e) & 0.7202&0.6037&0.6403 & 2 \\
		~&(2,e) &1.8985&2.3394&2.1821 & 3\\
		~&(3,e) &34.0832&30.8261&31.8757 & 1\\
		~&(1,h) &0.5956&0.9828&0.8317 & 1\\
		~&(2,h) &0.3014 &0.3669&0.3436 & 1\\ 
		\colrule
		\botrule
	\end{tabular} \\
	$m_0$ is free electron mass.
\label{EFF}
\end{table}

\subsection{Ce$_3$Te$_4$} 
\begin{figure}[htbp]
	\centering
	\includegraphics[width=6cm]{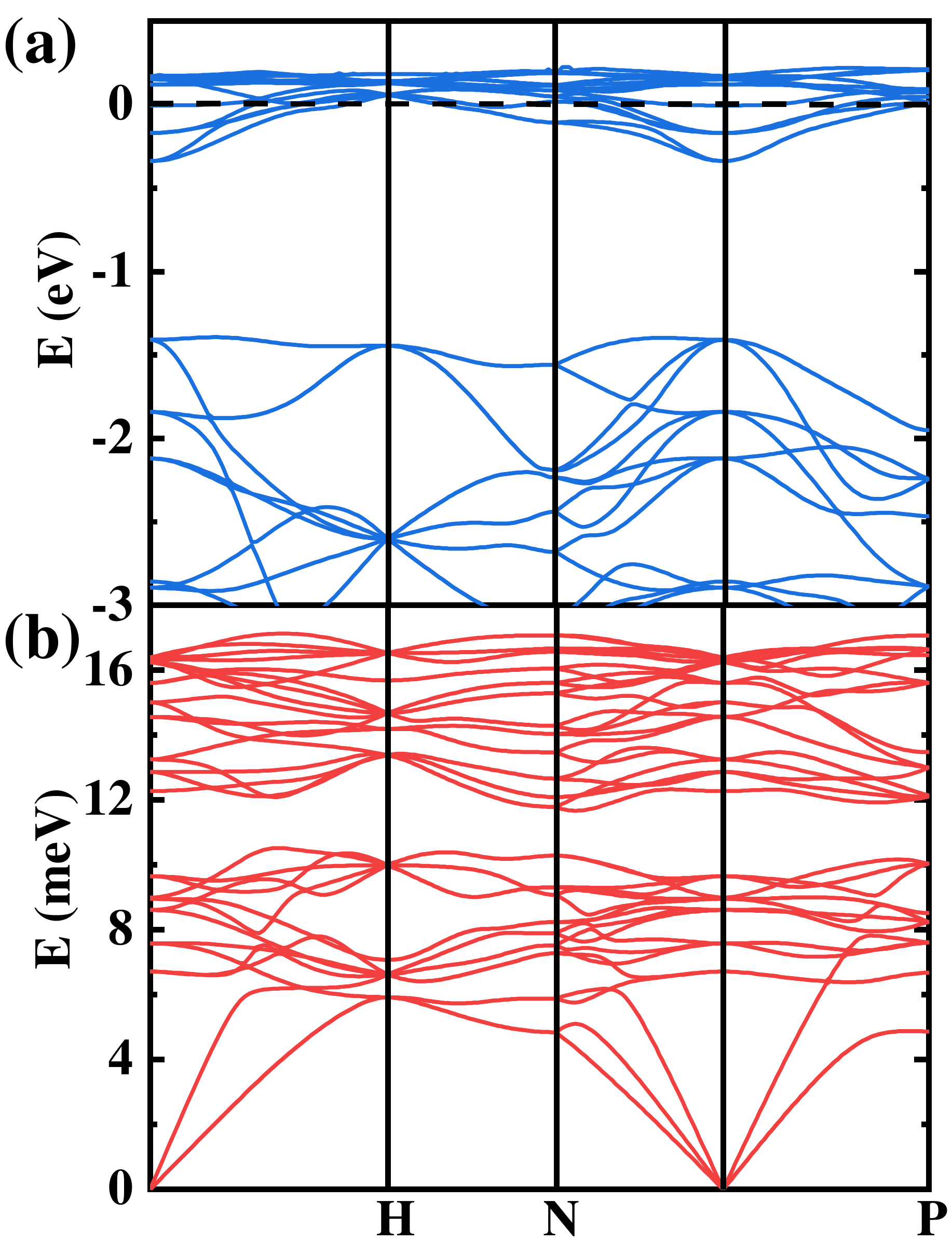}
	\caption{(a) The band structure and (b) phonon spectrum of  Ce$_3$Te$_4$.}
	\label{CETE-SP}
\end{figure}
Figure \ref{CETE-SP} show that the band structure and the phonon spectrum of Ce$_3$Te$_4$.  
Comparison with Figure \ref{LATE-SP}  shows that the energy band structures and phonon spectra of  Ce$_3$Te$_4$ and La$_3$Te$_4$ are similar, both are direct band gap semiconductor structures.
By comparing the data in Table \ref{PARA}, we can clearly find that the parameters of Ce$_3$Te$_4$ and La$_3$Te$_4$ are basically similar except for the large difference in lattice constants $a$ and optical mode energy $\textsc{A}_{2}$. 
At room temperature, electron-phonon scattering is weak and impurity scattering dominates the TE transport.
As the temperature increases, the lattice vibration gradually strengthens and electron-phonon scattering gradually dominates the TE transport.
And the energy value of $\textsc{A}_{2}$ has a large effect on the TE properties at high temperatures and no effect at room temperature.
This is the reason why Ce$_{3-x}$Te$_4$ and La$_{3-x}$Te$_4$ have similar TE properties at low temperatures. \cite{May2012}

From Table \ref{EFF}, it can be found that the effective mass of the nearest energy band (3, e) of Ce$_3$Te$_4$ near the Fermi energy level is much larger than the effective mass of the other energy bands. 
This is mainly because compared to La($5d6s^2$), Ce($4f5d6s^2$) has one more $4f$ electron which is localized, and this leads to a large effective mass of its bonding orbitals.
However, $4f$ electron does not contribute to the TE transport properties.\cite{shim2007modeling,li2022temperature}
Therefore, we can ignore the contribution of energy band (3, e) in the simulations.

\section{THERMOELECTRIC TRANSPORT PROPERTIES}
 
Three conduction bands should be considered in calculating the electron transport due to they are close enough. Besides, the bipolar transport should also be incorporated since holes will be excited and the electron-hole pair is formed in conduction band at high temperatures. 
The transport properties in Th$_3$Te$_4$ are calculated by the multiband BTE under the RTA.\cite{wang2011} Considering the TE properties of charge carriers in the lowest conduction band and the highest valence band, each of these bands is $N$-folded degeneracy, the dispersion relation of each carriers pocket can be expressed considering the nonparabolicity:
\begin{eqnarray}
\frac{\hbar^{2}k_{i,j,\|}^{2}}{2m^{\ast}_{i,j,\|}}+\frac{\hbar^{2}k_{i,j,\bot}^{2}}{2m^{\ast}_{i,j,\bot}}=\gamma(E_{i,j})=E_{i,j}+\frac{E_{i,j}^2}{E_g}
\end{eqnarray}
where $\hbar$ is the reduced Plank constant, $E_{i,j}$ is the energy, and $\gamma(E_{i,j})=E_{i,j}(1+E_{i,j}/E_{g})$.
The density-of-states effective mass of each band $m^{\ast}_{i,j,d}$ can be calculate by $m^{\ast}_{i,j,d}=N^{\frac{2}{3}}m^{\ast}_{i,j}$. 
For a fixed doping concentration ${n_d}$, the chemical potential $\mu$ in the  La$_3$Te$_4$ can be determined numerically.\cite{Zhou2010,wang2017theoretical} Assuming that all the scattering events are independent, the total relaxation time of each band ($\tau_{i,j}^{tot}$) can be expressed by the Mathiessen's rule:
\begin{eqnarray}
\frac{1}{\tau_{i,j}^{tot}}=\frac{1}{\tau^{imp}_{i,j}}+\frac{1}{\tau^{po}_{i,j,}}+\frac{1}{\tau^{da,l}_{i,j}}+\frac{1}{\tau^{do,l'}_{i,j}}
\end{eqnarray}
where $\tau^{imp}_{i,j}$ is the relaxation time of carries-impurity scattering, $\tau^{po}_{i,j}$ is that of carries-longitudinal polar optical phonon scattering, $\tau^{da,l}_{i,j}$ is that of carries-deformation acoustic phonon scattering corresponding to $l$th branch of acoustic phonon mode, and $\tau^{do,l'}_{i,j}$ is that of carries-deformation optical phonon scattering corresponding to $l'$th branch of optical phonon mode, respectively. 
In principle, the relaxation time for different scattering mechanisms can be obtained by Fermi’s golden rule.
The detailed temperature- and energy-dependent expressions for each scattering relaxation
time mentioned above can be found in Refs.\,\onlinecite{Zhou2010,nag2012electron}.

For bipolar transport, the TE transport coefficients such as electrical conductivity ($\sigma$), Seebeck coefficient (\emph{S}) and electronic thermal conductivity ($\kappa_{c}$) can be calculated by solving the BTE under the RTA. For anisotropic materials, the TE properties along different directions, which is denoted by $\xi=\|$ or $\bot$, can be written as, \cite{Zhou2010,wang2017theoretical}
\begin{eqnarray}
\sigma_{\xi} &=&\sum_{j}\sigma_{\xi,j},   \sigma_{\xi,j}=\sum_{j}\frac{q_{j}^{2}}{3\pi^{2}}(\frac{2k_{B}T}{\hbar^{2}})^{3/2}F_{0,\xi,j} \\
S_{\xi} &=&\sum_{j}\frac{S_{\xi,j}\sigma_{\xi,j}}{\sigma_{\xi}},   S_{\xi,j}=\frac{k_{B}}{q_{j}}(\frac{F_{1,\xi,j}}{F_{0,\xi,j}}-\eta_{j,\mu}) \\
\kappa_{c,\xi} &=&\sum_{j}\kappa_{c,\xi,j}+\frac{\sigma_{\xi,e}\sigma_{\xi,h}}{\sigma_{\xi}}(S_{\xi,e}-S_{\xi,h})^{2}T, \nonumber \\
\kappa_{c,\xi,j} &=&\frac{k_{B}^{2}T}{3\pi^{2}}(\frac{2k_{B}T}{\hbar^{2}})^{3/2}(F_{2,\xi,j}-\frac{F_{1,\xi,j}^{2}}{F_{0,\xi,j}}) \\
F_{n,\xi,j}&=&\sum_{i}\frac{m^{\ast 3/2}_{i,j,d}}{m^{\ast}_{i,j,\xi}}\int_{0}^{\infty}\eta_{i,j}^{n}\gamma^{\frac{3}{2}}(\eta_{i,j})\tau_{i,j}^{tot}(-\frac{\partial f_{0}}{\partial \eta_{i,j}})d\eta_{i,j} \label{F-EQ}
\end{eqnarray}
where $q_{j}$ denotes the charge of carriers, $k_{B}$ is the Boltzmann constant, $\eta_{i,j}=\frac{E_{i,j}}{k_{B}T}$, $\eta_{e,\mu}=\frac{\mu-E_{g}}{k_{B}T}$, $\eta_{h,\mu}=-\frac{\mu}{k_{B}T}$, $\eta_{g}=\frac{E_{g}}{k_{B}T}$, and $\gamma(\eta_{i,j})=\eta_{i,j}(1+\frac{\eta_{i,j}}{\eta_{g}})$, respectively. $f_{0}$ in Eq. (\ref{F-EQ}) is the equilibrium Fermi-Dirac distribution. 

\subsection{TE properties of La$_3$Te$_4$} 

\begin{figure}[htbp]
	\centering
	\includegraphics[width=8cm]{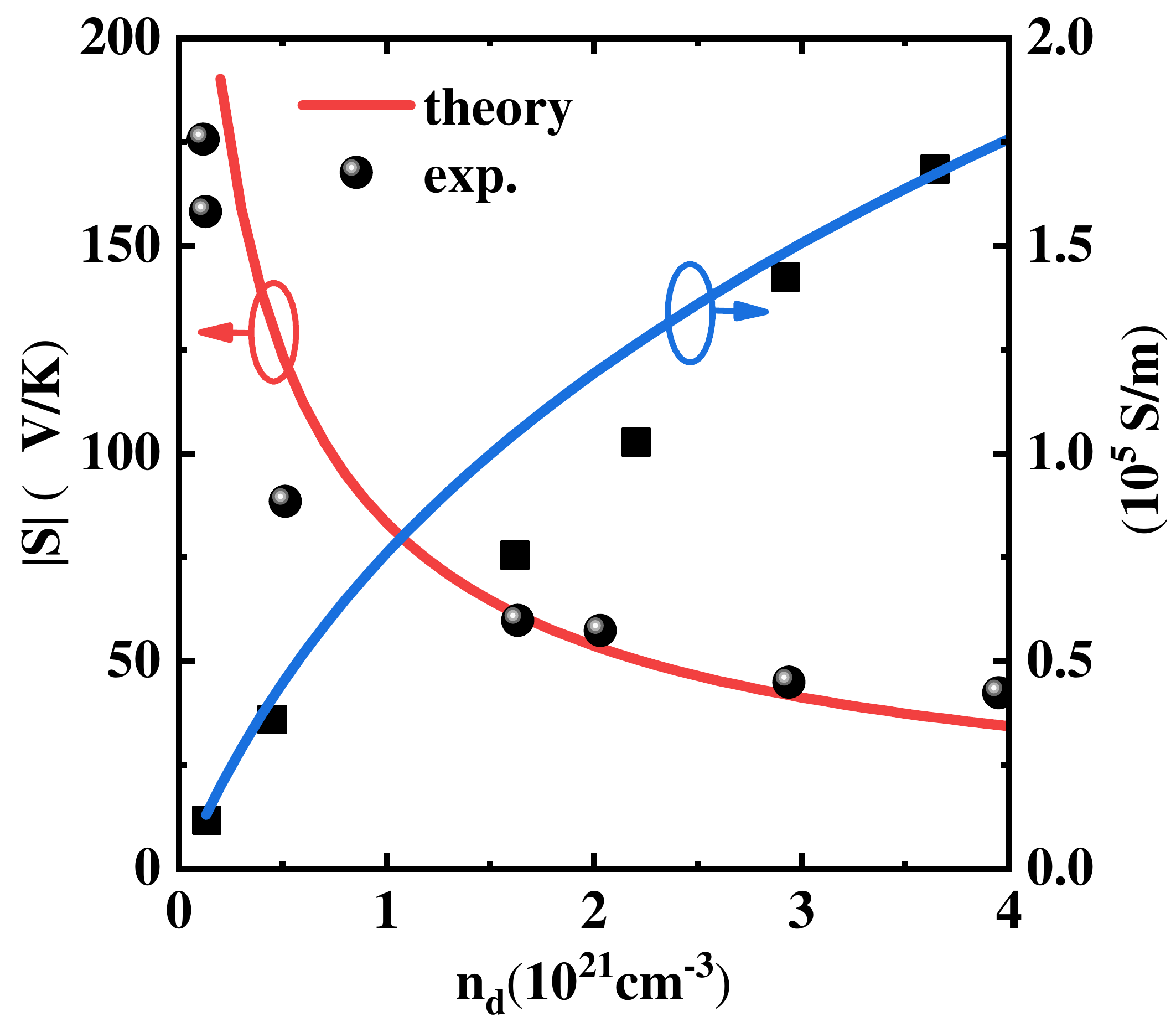}
	\caption{Calculated Seebeck coefficient (a) and electrical conductivity (b) of La$_{3}$Te$_4$ as a function of carrier concentration. The experimental data are extracted from Ref.\,\onlinecite{May2009}.}
	\label{LATE-FITTING1}
\end{figure}

We now turn to calculate the electrical conductivity and the Seebeck coefficient of La$_3$Te$_4$ based on the band structures of La$_3$Te$_4$ obtained from first-principles calculations in Sec.\ref{Band} A. In order to justify the input parameters in our calculation, we first fit the experimental data of La$_3$Te$_4$ reported by Ref.\,\onlinecite{May2009}. The isotropic electrical conductivity along different directions is averaged to compare to the measured electrical conductivity. Figure \ref{LATE-FITTING1} shows that the calculated electrical conductivity and the Seebeck coefficient as a function of carrier concentration are in good agreement with the experimental results. Table \ref{LA-FIT} presents the reasonable fitted parameters adopted in our calculations. An increase of electrical conductivity and a decrease of the Seebeck coefficient with increasing carrier concentration comes from the $\sigma \propto n_{d}$ and $S \propto n_{d}^{-2/3}$. 

\begin{table}
	\caption{Fitting parameters used to calculate the transport coefficients in La$_3$Te$_4$ at 400K.}
	\begin{tabular}{@{}lcc}
		\toprule
	    Parameters & Fitted value \\
	    \colrule
        mass density ($\text{g cm}^{-3}$) & 6.92\cite{goodenough1970magnetic} \\
        optical phonon energy (meV) & 6.2\cite{Viennois2013} \\
        deformation potential constant (eV) & 6.1\cite{May2010} \\
        high-frequency permittivity ($\epsilon_{0}$) & 2.7 \\
        static permittivity ($\epsilon_{0}$) & 27\cite{May2010}\\
        impurity density ($10^{19}\text{cm}^{-3}$) & 8\cite{May2012} \\
		\colrule
		\botrule
	\end{tabular} \\
\label{LA-FIT}
\end{table}

\begin{figure}[htbp]
	\centering
	\includegraphics[width=8cm]{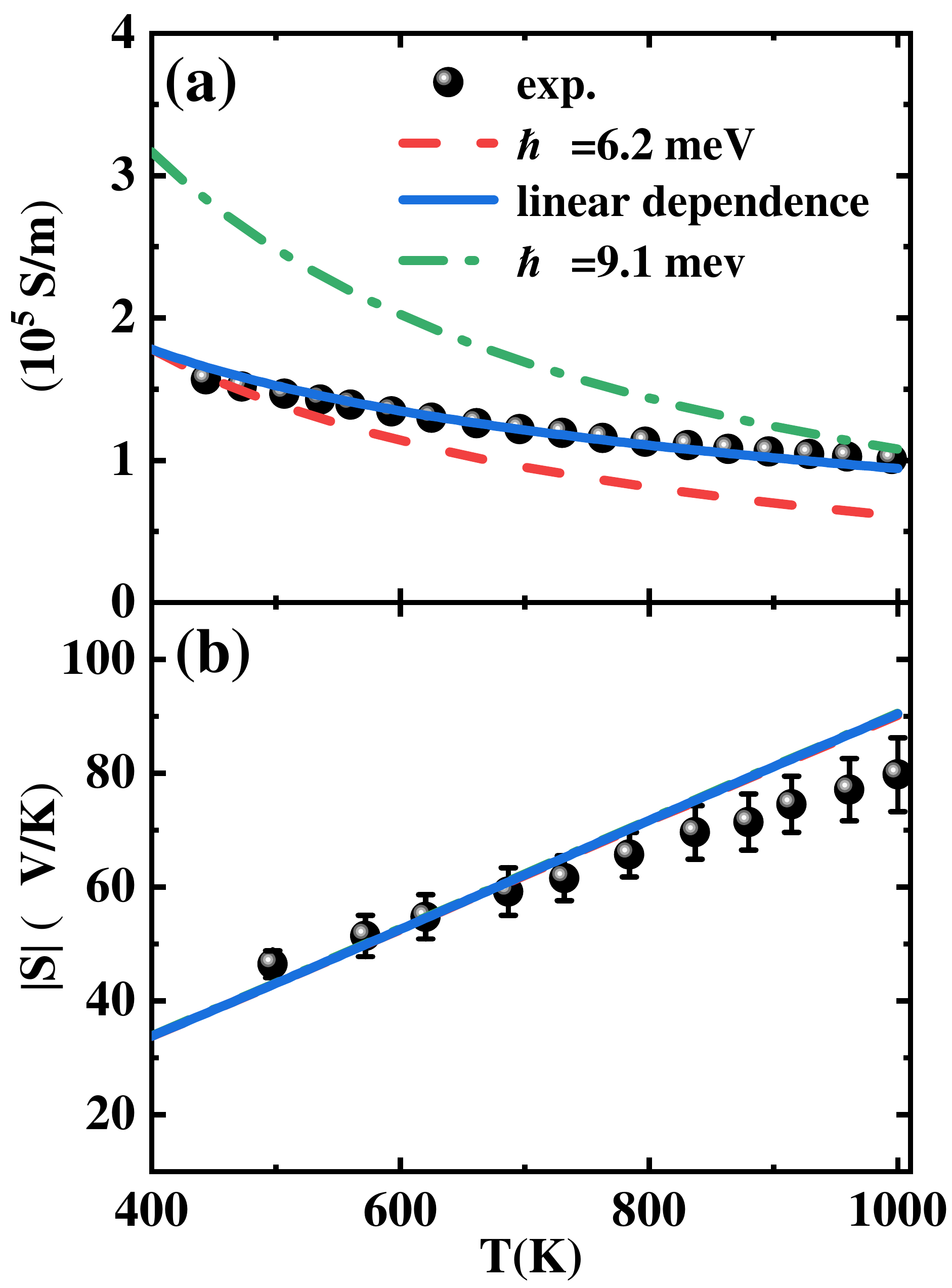}
	\caption{Calculated Seebeck coefficient (a) and electrical conductivity (b) of La$_{3}$Te$_4$ as a function of temperature. The constant optical branch phonon energies (red dashed and green dotted lines) do not describe well their electrical conductivity experimental data. For the linear dependence, the optical branched phonon energy fits well with the experimental values of electrical conductivity. For the Seebeck coefficients, the fit results remain largely consistent for the three different modes. The experimental data are extracted from Ref.\onlinecite{May2008}.}
	\label{LATE-FITTING2}
\end{figure}

On this basis, we can further investigate the TE properties of La$_3$Te$_4$ as a function of temperature, as shown in Figure \ref{LATE-FITTING2}.
At low temperatures, impurity scattering is dominant for the TE properties and other scattering is less influential. As the temperature increases, the lattice vibration becomes strong, intensifying electron-phonon scattering. In this case, the electron-deformation acoustic (optical) phonon scattering will dominate the TE properties. Following the Einstein model to approximate all optical branches as constant frequencies, the numerical simulation results do not fit the experiment well, as shown by the red dashed line and the green dotted line in Figure \ref{LATE-FITTING2} (a).
When the optical phonon frequency is small, the numerical simulation value is smaller than the experimental value in the high temperature region. Conversely, when the optical phonon frequency is larger, the theoretical value will gradually approach the experimental value as the temperature increases.
This is due to the fact that, like phonon-phonon scattering, electron-phonon scattering also has normal scattering (N process) and inversion scattering (U process).
According to the Debye model, the phonon energy (meV) is about a few thousandths of the electron energy on the Fermi surface. 
Therefore, the change of the electron energy is almost negligible due to electron-phonon scattering. Although electron-phonon scattering can be considered as completely elastic scattering, it changes the direction of electron motion, which has a significant effect on electrical conductivity.
At low temperatures, the electron scattering angle is small because only low-frequency phonon modes are excited, which has limited effect on the conductivity. As the temperature rises, more vibrational modes of phonons will be excited. 
The angle of phonon and electron scattering differs for different modes, which will also have different effects on the conductivity.
Here, for simplicity, we assume a linear dependence of the energy of the different modes of phonons scattered with electrons on temperature as, $\hbar \omega = \hbar \omega_{0} + 3.412\times10^{-3}(T-T_{0})$, where $\hbar \omega_{0} = 6.2$meV, $T_{0} =400$K.
The calculated results using linear dependence are in good agreement with the experimental values, as shown by the blue solid line in Figure \ref{LATE-FITTING2}(a).
The Seebeck coefficient is mainly measured by the average energy magnitude of carriers, which is related to the density of states near the Fermi surface. And the effect of electron-phonon scattering for the average carrier energy can be neglected. Therefore, the Seebeck coefficients fitted by the three different methods are essentially comparable, as shown in the Figure \ref{LATE-FITTING2}(b).

\subsection{TE properties of Ce$_3$Te$_4$} 

\begin{figure}[htbp]
	\centering
	\includegraphics[width=8cm]{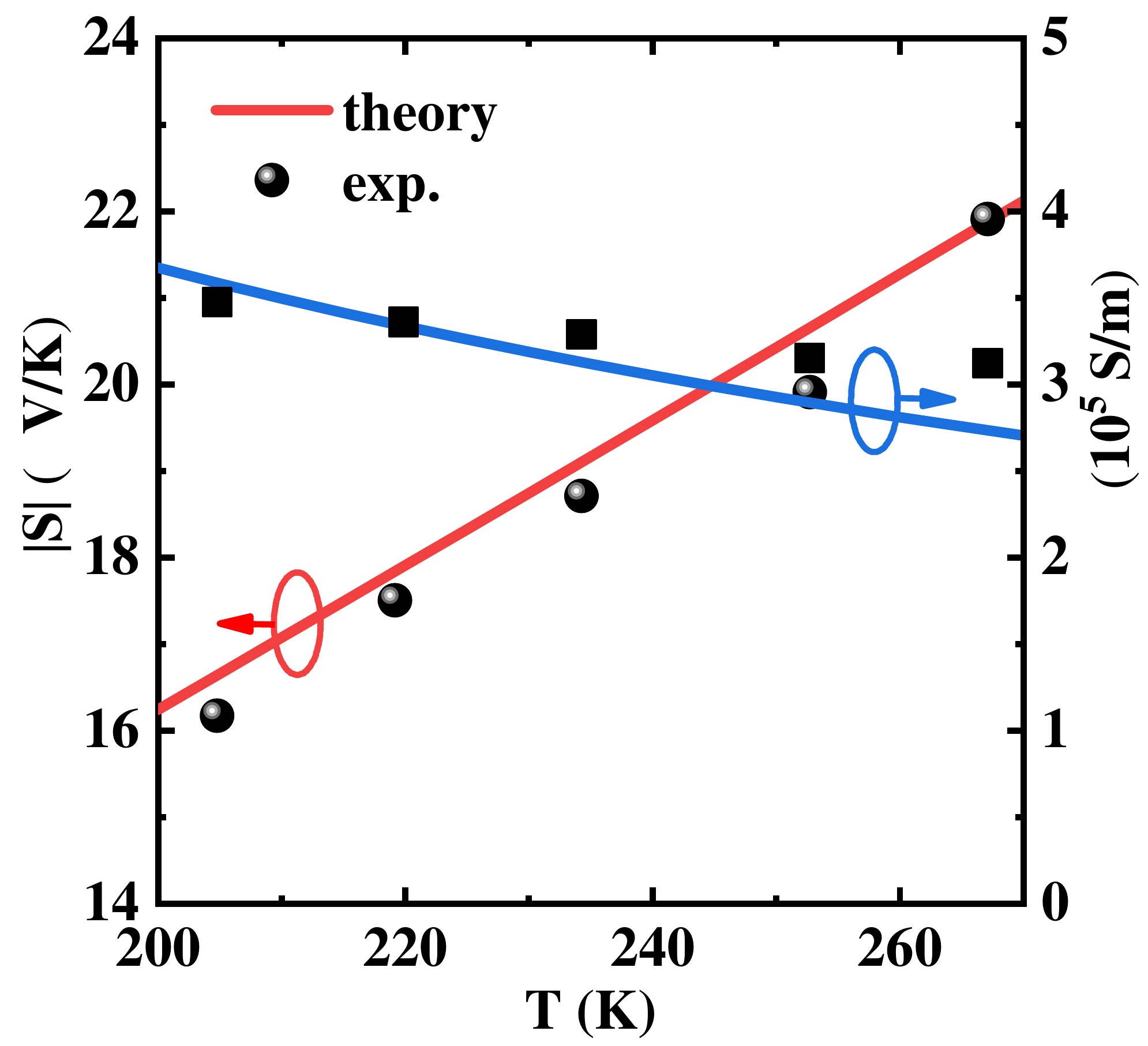}
	\caption{Calculated Seebeck coefficient (a) and electrical conductivity (b) of Ce$_{3}$Te$_4$ as a function of temperature. The experimental data are extracted from Ref.\,\onlinecite{May2012}.}
	\label{CETE-FITTING}
\end{figure}

Considering that the local state electrons do not contribute to the electron transport, we only consider the contribution of the five lowest conduction bands and the two highest valence bands to the TE transport properties of Ce$_3$Te$_4$.
Figure \ref{CETE-FITTING} depicts the calculated electrical conductivity and Seebeck coefficient versus temperature based on the BTE under the RTA.
The fitting parameters used in the calculations are shown in Table \ref{CE-FIT}. We can find that the calculated values are consistent with the experimental data, which indicates that the fitting parameters are chosen reasonably.
In addition, it shows a decrease in electrical conductivity and an increase of the Seebeck coefficient with increasing temperature.
This is mainly due to the increase of carrier scattering intensity at higher temperatures and the increase of the average energy carried by carriers.

\begin{table}
	\caption{Fitting parameters used to calculate the transport coefficients in Ce$_3$Te$_4$ at 400K.}
	\begin{tabular}{@{}lcc}
		\toprule
	    Parameters & Fitted value \\
	    \colrule
	    carriers concentration ($10^{21}\text{cm}^{-3}$) & 4.6\cite{May2012,Vo2014}\\
        mass density ($\text{g cm}^{-3}$) & 7.12\cite{goodenough1970magnetic} \\
        optical phonon energy (meV) & 6.2 \\
        deformation potential constant (eV) & 6.1 \\
        high-frequency permittivity ($\epsilon_{0}$) & 2.7 \\
        static permittivity ($\epsilon_{0}$) & 27\\
        impurity density ($10^{19}\text{cm}^{-3}$) & 5 \cite{May2012}\\
		\colrule
		\botrule
	\end{tabular} \\
\label{CE-FIT}
\end{table}

\subsection{Comparison of TE properties of La$_3$Te$_4$ and Ce$_3$Te$_4$} 

\begin{figure}[htbp]
	\centering
	\includegraphics[width=8cm]{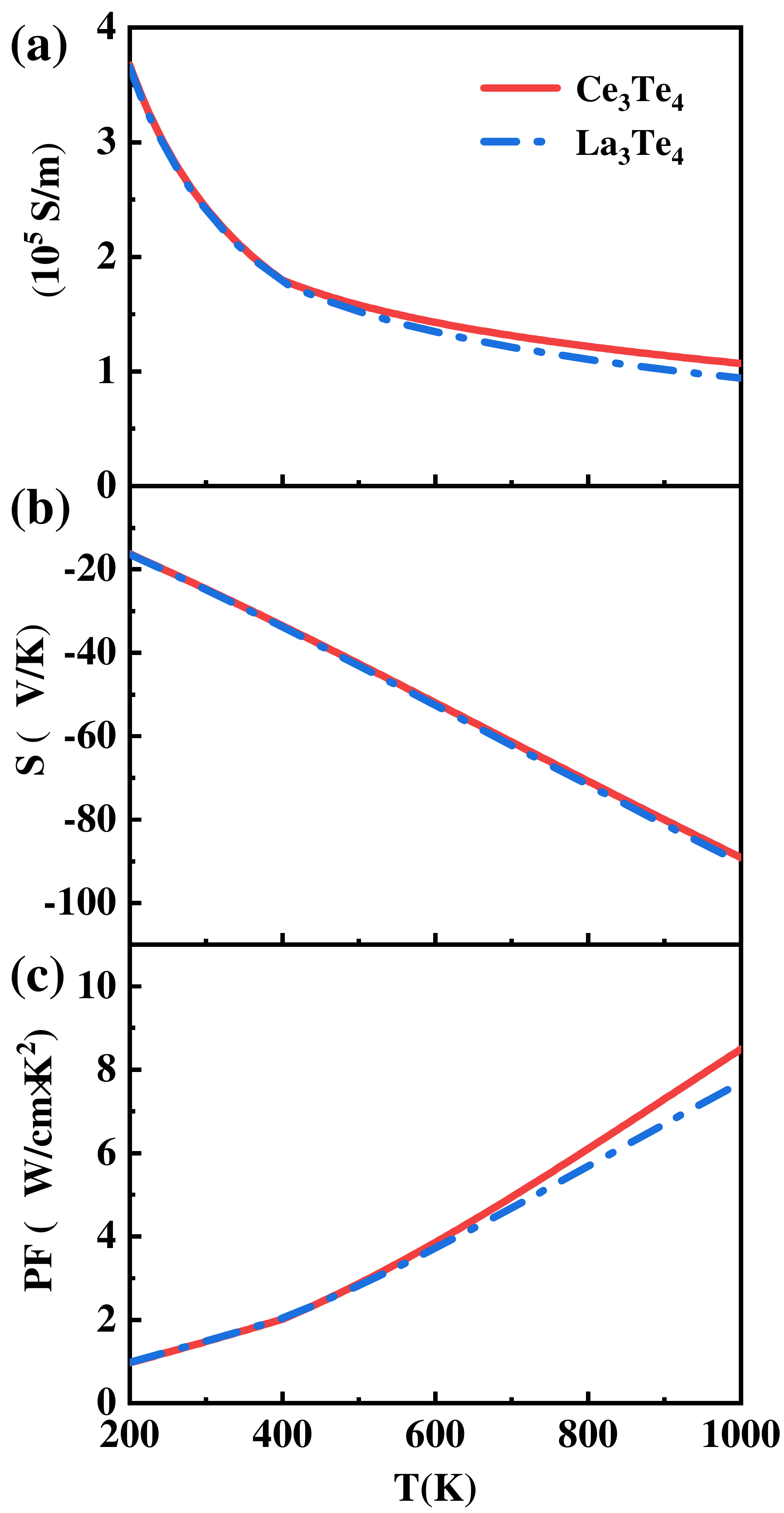}
	\caption{Comparison of TE properties of La$_3$Te$_4$ and Ce$_3$Te$_4$: Electrical conductivity (a), Seebeck coefficient (b), and Power factor (c) as a function of temperature.}
	\label{COMP}
\end{figure}

At room temperature, experimental and theoretical reports have shown that the Ce$_3$Te$_4$ has similar TE transport properties as La$_3$Te$_4$, such as electrical conductivity, Seebeck coefficient, and power factor. \cite{May2008,May2012}
We can obtain the same conclusion by numerical simulation using BTE under the RTA as shown in Figure \ref{COMP}.
Since the effective mass of Ce$_3$Te$_4$ is larger than that of La$_3$Te$_4$, the electrical conductivity of Ce$_3$Te$_4$ is reasonably smaller than that of La$_3$Te$_4$. However, at low temperatures, the electrical conductivity of La$_3$Te$_4$ and Ce$_3$Te$_4$ are approximately equal, which indicates that the average relaxation time of Ce$_3$Te$_4$ is larger than that of La$_3$Te$_4$.
The effect of scattering will be greater with increasing of temperature. In the high temperature region, more phonon modes are excited, which leads to a rise in the number of phonons and allows phonons to  participate in TE transport.
At this point, electron-phonon scattering, which consists of the polar optical phonons, the deformed acoustic phonons, and the deformed optical phonons, dominates TE transport.
As shown in Figure \ref{COMP}(a) and (b), the Seebeck coefficient of Ce$_3$Te$_4$ is comparable to that of La$_3$Te$_4$ at high temperatures, but its electrical conductivity is higher than that of La$_3$Te$_4$.
For example, the electrical conductivity of Ce$_3$Te$_4$ and La$_3$Te$_4$ at $T = 1000$K are 1.07$\times10^{5}$ and 0.942$\times10^{5}$ S/m, respectively.
This is mainly due to the difference in the optical branch phonon energy of A$_{2}$ mode of La$_3$Te$_4$ and Ce$_3$Te$_4$, as shown in Table \ref{PARA}.
From the phonon spectrum, we find that the optical branch phonon energy of A$_{2}$ mode of Ce$_3$Te$_4$ is slightly larger than that of La$_3$Te$_4$.
The magnitude of the optical branch phonon energy will determine the magnitude of relaxation time of the electron-deformation optical phonon scattering. 
A larger energy will correspond to a larger relaxation time.
In the case of high temperatures, electron-phonon scattering plays a dominant role, which has increased the total relaxation time.
Therefore, the electrical conductivity of Ce$_3$Te$_4$ is larger than that of La$_3$Te$_4$ under other equal conditions.
Moreover, the power factor of Ce$_3$Te$_4$ is also better than that of La$_3$Te$_4$ at high temperature.

\subsection{Optimal carrier concentration of Ce$_3$Te$_4$ at various temperatures} 

\begin{figure}[htbp]
	\centering
	\includegraphics[width=8cm]{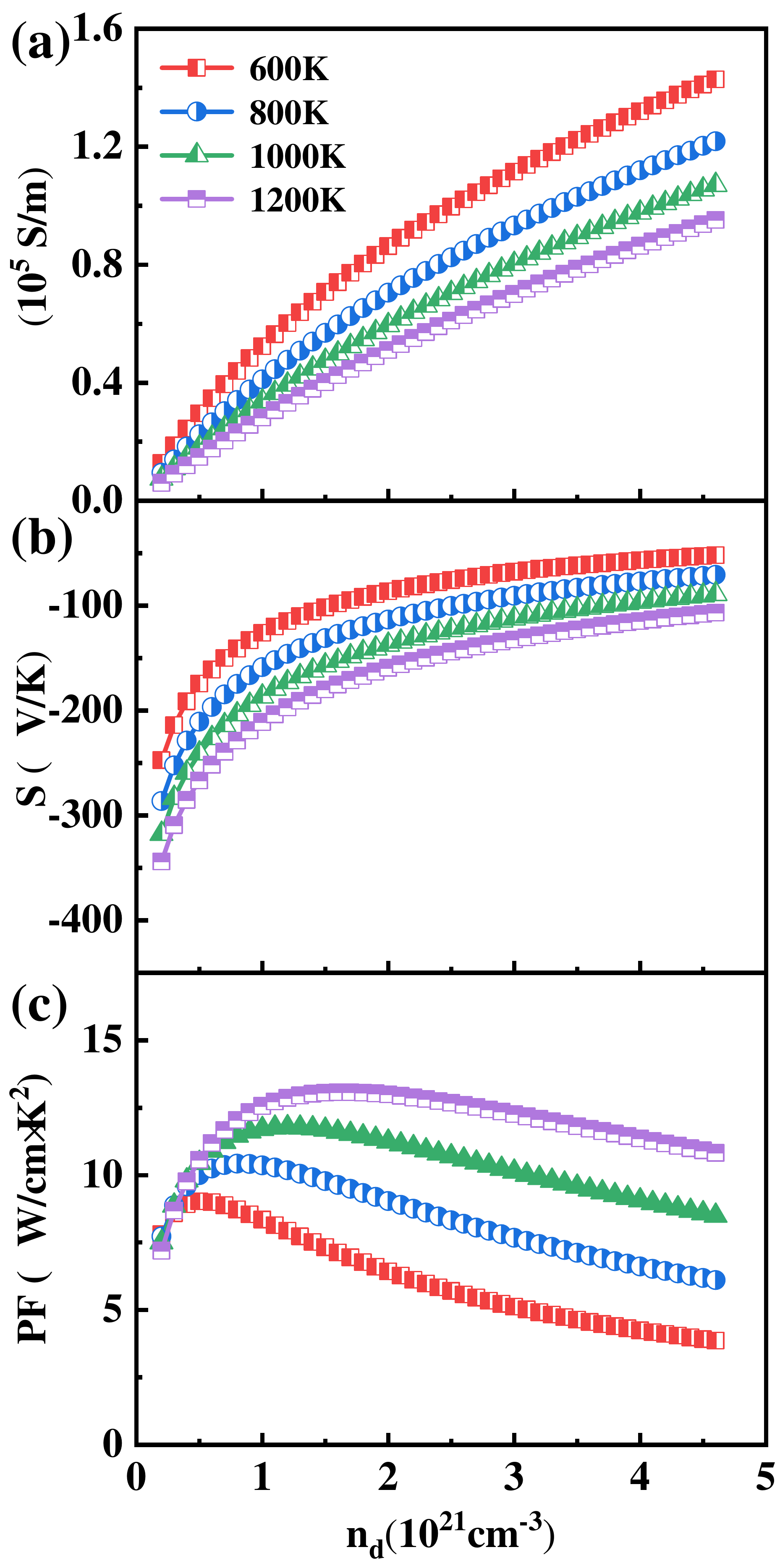}
	\caption{Calculated TE properties of Ce$_3$Te$_4$: Electrical conductivity (a), Seebeck coefficient (b), and Power factor (c) as a function of carrier concentration at various temperatures.}
	\label{OPT}
\end{figure}

To further investigate the TE properties of Ce$_3$Te$_4$ at high temperatures, we will consider the TE transport properties of Ce$_3$Te$_4$ as a function of carrier concentration with various high temperatures, as shown in Figure \ref{OPT}.
Due to $\sigma \propto n_{d}$ and $\sigma \propto \tau_{tot} \propto 1/T$, the electrical conductivity will be enhanced when the carrier concentration increases or the temperature decreases as shown in Figure \ref{OPT}(a).
Due to the coupling relationship between $\sigma$ and $S$, the Seebeck coefficient changes in the opposite direction.
However, it can be seen from Figure \ref{OPT}(b) that the trend of Seebeck variation becomes gradually smoother as the carrier concentration increases.
This is because the bipolar effect will be more pronounced at high temperatures by exciting the intrinsic carrier.
In particular, fora lower carrier concentration, the bipolar effect has a greater impact on the Seebeck coefficient, which can be found to decrease at a greater rate than that at higher carrier concentrations.
Figure \ref{OPT}(c) shows the dependence of the power factor on the carrier concentration at different temperatures.
We found that the power factor corresponding to the higher temperature is smaller as the carrier concentration is low.
Because of the inhibitory effect of minority carriers on the TE properties, although the intrinsic excitation increases the minority carrier concentration and the electrical conductivity, the presence of the minority carriers can cause a decrease in the Seebeck coefficient, which is more than compensate for the increase in the Seebeck coefficient caused by the increase in conductivity.
As the carrier concentration increases, the effect of the bipolar diffusion effect on the TE performance generated by the minority carriers diminishes. In other words, the bipolar effect can be suppressed via the heavily doped method.
In addition, the optimal carrier concentration of Ce$_3$Te$_4$ varies for different temperatures, and the corresponding power factors are also different.
For example, the optimal carrier concentration is around $0.5\times10^{21}$cm$^{-3}$ with the peak power factor 9.02 $\mu$Wcm$^{-1}$K$^{-2}$ at $T=600$K; the optimal carrier concentration is around $0.8\times10^{21}$cm$^{-3}$ with the peak power factor 10.42 $\mu$Wcm$^{-1}$K$^{-2}$ at $T=800$K; the optimal carrier concentration is around $1.2\times10^{21}$cm$^{-3}$ with the peak power factor 11.78 $\mu$Wcm$^{-1}$K$^{-2}$ at $T=1000$K; And the optimal carrier concentration is around $1.6\times10^{21}$cm$^{-3}$ with the peak power factor 13.07 $\mu$Wcm$^{-1}$K$^{-2}$ at $T=1200$K.
This is because the carrier concentration of the intrinsic excitation is strongly correlated with the temperature, and as the temperature increases, the optimal carrier concentration shifts upward.

\section{CONCLUSION}
We have incorporated the multiband Boltzmann transport equations with first-principles calculations on electronic band structures in order to theoretically investigate TE properties of Th$_3$Te$_4$ materials such as La$_3$Te$_4$ and Ce$_3$Te$_4$.
Our theoretical calculations are in good agreement with the experimental data with calculated parameters and several other fitting parameters.
Theoretical results show that for the TE transport properties at high temperatures, a linear dependence is more consistent with the experimental results than the constant optical branch phonon energy describing the electron-deformation optical branch phonon scattering.
In addition, we predict the TE transport properties of Ce$_3$Te$_4$ at high temperatures and the optimal carrier concentration at different temperatures, which is a guideline for experimental aspects.

\begin{acknowledgments}
This work was supported by National Key R{\&}D Program of China
(No. 2017YFB0406004), National Natural Science Foundation of China
(No. 11890703), Key-Area Research and Development Program of Guangdong Province (No. 2020B010190004).
\end{acknowledgments}

\bibliography{CeTe}
% Encoding: UTF-8

\end{document}